\documentstyle[psfig]{caosp}

\begin{document}

\htitle{Search for the $^{3}$He isotope...}
\hauthor{I. Stateva {\it et al.}}

\title{Search for the $^{3}$He isotope in the atmospheres of HgMn stars}

\author{I. Stateva \inst{1} \and T. Ryabchikova \inst{2}
 \and I. Iliev \inst{3}}

\institute{Institute of Astronomy,
           blvd. Tzarigradsko shoussee 72, Sofia 1784, Bulgaria \\
 \and Institute of Astronomy,
     Pyatnitskaya 48, 109017 Moscow, Russia \\
 \and NAO - Rozhen, P.O.Box 136,
     4700 Smolyan, Bulgaria \\
}
\maketitle

\section{Introduction}

For the last twenty years it has been established that the $^{3}$He isotope
is present in the spectra of some peculiar stars, which ocupy a narrow
range of effective temperatures. An explanation of these results has been
attempted in the framework of the diffusion theory (Michaud et al.,
1979). On the other hand, the He\,{\sc i} lines used for detecting the
$^{3}$He isotope are burdened with blends, which were unknown at the times
of the first studies of $^{3}$He; therefore, careful reanalysis of the
$^{3}$He isotope is needed before making comparisons with the predictions
of helium diffusion .

Three of the stars studied here were checked for the presence of $^{3}$He
by Hartoog \& Cowley(1979). They searched for $^{3}$He on photographic
spectra by accurately measuring the isotopic shifts of the He\,I lines.
We performed our work as a continuation of a systematic study of the
$^{3}$He isotope in stellar atmospheres, on the basis of high dispersion,
high S/N spectra and using the spectral synthesis technique.

\section{Observations and results}

We carried out observations of HD\,58661, HD\,172044, HD\,185330 and
HD\,186122 (46\,Aql) at the Bulgarian National Astronomical Observatory
(BNAO) - Rozhen with the CCD camera attached to the Coud\'e-spectrograph of
the 2m telescope. The CCD spectra were centred on three regions around
He\,{\sc i} lines ($\lambda\lambda$ 4921, 5875, 6678). The spectral resolving
power was about 30\,000 and the signal-to-noise ratio (S/N) was 150-200.

Atmosphere parameters $T_{\rm eff}$ and $\log g$ were derived using
photometric data in Geneva and Str\"omgren photometric systems. The
microturbulences were taken from Smith \& Dworetsky(1993). For HD\,185330
we assumed $\xi$=0 km~s$^{-1}$.

\begin{table}
\small
\caption{Model atmosphere parameters, helium abundances and isotopic ratios}
\label{t1}
\begin{center}
\begin{tabular}{ccccccc}
\hline
star & $T_{\scriptstyle eff}$ & $\log g$ & $\xi$ &
$v_{\scriptstyle e} \sin i$ & $\log{\rm He/H}$ & $^{3}$He/$^{4}$He \\
 & [K] & [dex cm s$^{-2}$] & [km/s] & [km/s] & & \\
\hline
HD\,58661  & 13\,500 & 4.0 & 0.5 & 30 & -1.70 & 0.56 \\
HD\,172044 & 14\,500 & 3.9 & 1.5 & 34 & -1.85 & \\
HD\,185330 & 16\,500 & 3.7 & 0.0 &  8 & -2.20 & 0.96 \\
46\,Aql    & 13\,000 & 3.7 & 0.0 & 15 & -2.55 & 0.1 \\
\hline
\end{tabular}
\end{center}
\end{table}

We used the spectral synthesis technique for checking the presence of the
$^{3}$He isotope (SYNTH code written by Piskunov 1992). The parameters of
the atomic lines  were extracted from the VALD database (Piskunov et al.
1995). In order to determine accurately the abundance of the $^{3}$He
isotope, we had to take into account the possible blends. All He\,{\sc i}
lines used are weak and the contribution of blends could be significant.
The He\,{\sc i} $\lambda$ 4921.93 \AA\, line is blended only with
Fe\,{\sc ii} $\lambda$ 4922.19 \AA, while He\,{\sc i} $\lambda$
6678.15 \AA\, may be blended with Ne\,{\sc i} $\lambda$ 6678.28 \AA,
Si\,{\sc ii} $\lambda$ 6678.66 \AA, Si\,{\sc ii} $\lambda$ 6678.73 \AA,
Fe\,{\sc ii} $\lambda$ 6678.84 \AA.

\begin{figure}
\psfig{figure=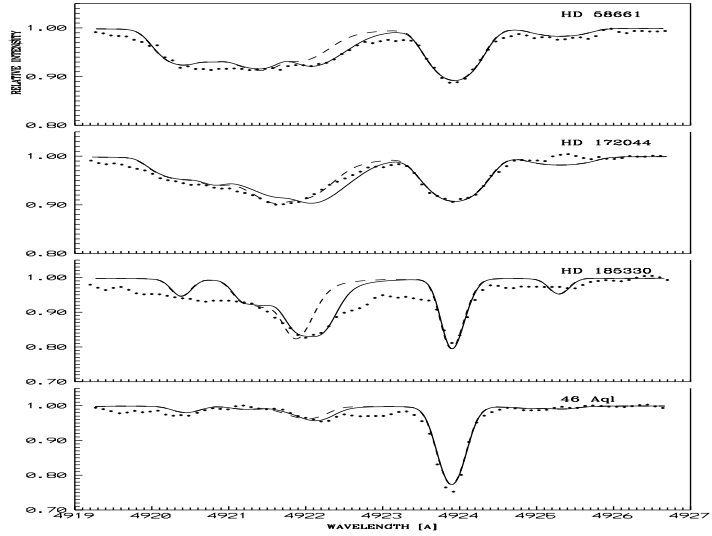,width=12cm,height=8cm}
\caption{Comparison of the observed (asteriks) and synthesized with
the presence of $^{3}$He isotope (full line) and without it (dashed
line) spectra for the region of $\lambda$ 4921}
\label{f1}
\end{figure}

\begin{figure}
\psfig{figure=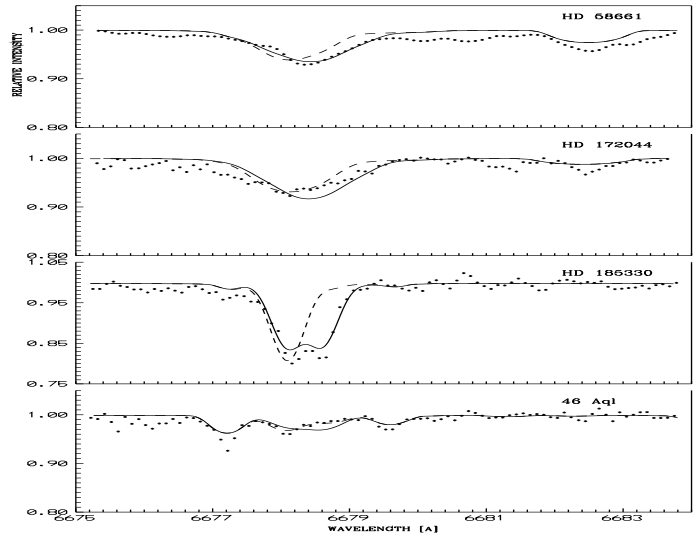,width=12cm,height=8cm}
\caption{The same as in Fig.1 but for the region of $\lambda$ 6678}
\label{f2}
\end{figure}

\underline{HD\,58661.}
We detected the presence of the $^{3}$He isotope in the atmosphere of
HD\,58661. He\,I $\lambda$ 4921 is blended with Fe\,II $\lambda$ 4922.19
\AA, which position nearly coincides with those of $^{3}$He\,I $\lambda$
4922.26 \AA. Because of the underabundance of Fe, this line could not
produce a significant amount of absorption (Figs~\ref{f1}, ~\ref{f2}).

\underline{HD\,172044.}
The line profiles of He\,I $\lambda$ 4921 and $\lambda$ 6678 were
synthesized successfully without any contribution of $^{3}$He. Our results
confirmed the conclusion of Hartoog \& Cowley(1979) that HD\,172044 is
not a $^{3}$He star.

\underline{HD\,185330.}
The presence of the $^{3}$He isotope in the atmosphere of HD\,185330 was
confirmed as it was reported for the first time by Hartoog \& Cowley(1979).
The isotope ratio $^{3}\rm He/^{4}\rm He$=0.96 derived by us does not
differ very much from that given by Hartoog \& Cowley, which was
$^{3}\rm He/^{4}\rm He$= 1.3.

\underline{46\,Aql.}
The $^{3}$He isotope was detected in the profile of $\lambda$ 4921
(Fig.~\ref{f1}) based on three spectra. But $\lambda$ 6678 was
synthesized more successfully without any presence of $^{3}$He isotope
based on one spectrum (Fig.~\ref{f2}). The reason for this discrepancy
may be found in the weakness of this line; moreover the helium
abundance is extremely low: $\log(\rm He/H)$=-2.55. 46\,Aql is a possible
$^{3}$He star.

\acknowledgements
{We are very grateful to Dr. F. Kupka (Institute of Astronomy, Vienna), who
helped us in calculating Kurucz model atmospheres. This work was
supported by the Bulgarian Ministry of SEC (grant F-603/96).}

\end{document}